\documentclass{appolb}
\usepackage{epsfig}

\begin{document}

\title{Composition of fluctuations of different observables}
\author{Grzegorz Wilk\thanks{e-mail: wilk@fuw.edu.pl}
\address{The Andrzej So{\l}tan Institute for Nuclear Studies,
Ho\.{z}a 69, 00681, Warsaw, Poland} \and Zbigniew W\l
odarczyk\thanks{e-mail:zbigniew.wlodarczyk@ujk.kielce.pl}
\address{Institute of Physics,Jan Kochanowski University,
\'Swi\c{e}tokrzyska 15, 25-406 Kielce, Poland} \and Wojciech
Wolak\thanks{e-mail: fizww@tu.kielce.pl}
\address{Kielce University of Technology, Tysi\c{a}clecia Pa\'nstwa
Polskiego 7, 25-314 Kielce Poland} }

\maketitle

\begin{abstract}

We demonstrate that description of fluctuations observed in
multiparticle production processes using Tsallis statistics
approach (in which fluctuations are described by the
nonextensivity parameter $q$) leads to a specific sum rule for
parameters $q$ seen in different observables which can be verified
experimentally.
\end{abstract}


\vspace{1cm}

\section{Introduction}
\label{section:I}

When analyzing multiparticle production data the standard tool
used is statistical modelling \cite{MG_rev}. However, this
approach does not account for the possible intrinsic
nonstatistical fluctuations in the hadronizing system which
usually result in a characteristic power-like behavior of the
single particle spectra and in the broadening of the corresponding
multiplicity distributions (and which can signal a possible phase
transition(s) \cite{PhTr}). One possibility to include this and
still remain in the domain of a statistical approach is to use the
so called Tsallis statistics \cite{T,WW_epja,BPU_epja}
(represented by Tsallis distribution, $h_q(E)$) which accounts for
such situations by introducing in addition to the temperature $T$,
a new parameter, $q > 1$, directly connected to fluctuations
\cite{WW,BJ} (for $q \rightarrow 1$ one recovers the usual
Boltzmann-Gibbs distribution, $f(E)$):
\begin{eqnarray}
h_q(E) &=& \frac{2-q}{T}\exp_q \left(-\frac{E}{T}\right) =
\frac{2-q}{T}\left[1 - (1-q)\frac{E}{T}\right]^{\frac{1}{1-q}}
\label{eq:Tsallis}\\
&& \stackrel{q \rightarrow 1}{\Longrightarrow}\, \frac{1}{T}\exp
\left(-\frac{E}{T}\right),\label{eq:BG}
\end{eqnarray}
The most recent applications of this approach come from the PHENIX
Collaboration at RHIC \cite{PHENIX} and from the CMS Collaboration
at LHC \cite{LHC_CMS}.  The parameter $q$ is entirely given by
intrinsic fluctuations in the system, cf. Eq. (\ref{eq:defq})
below\footnote{See also \cite{CTB}. One must admit at this point
that this approach is subjected to a rather hot debate of whether
it is consistent with equilibrium thermodynamics or else it is
only a handy way to phenomenologically describe some intrinsic
fluctuations in the system \cite{debate}. However, as was recently
demonstrated on general grounds in \cite{M}, fluctuation phenomena
can be incorporated into a traditional presentation of
thermodynamic and Tsallis distribution, Eq. (\ref{eq:Tsallis}),
belongs to the class of general admissible distributions which
satisfy thermodynamical consistency conditions and which are
therefore a natural extension of the usual Boltzman-Gibbs
canonical distribution Eq. (\ref{eq:BG}).}.\\

Before proceeding any further we must emphasize two points. First,
the relation between parameter $q$ and fluctuation of temperature
was derived in \cite{WW}, where it was shown that starting from
some simple diffusion picture of temperature equalization in a
nonhomogeneous heat bath (in which local $\tilde{T}$ fluctuates
from point to point around some equilibrium temperature, $T$) one
obtains an evolution of $\tilde{T}$ in the form Langevin
stochastic equation and distribution of $1/\tilde{T}$,
$f(1/\tilde{T})$, as solution of the corresponding Fokker-Planck
equation. It turns out that $f(1/\tilde{T})$ has the form of a
gamma distribution,
\begin{eqnarray}
f(1/\tilde{T}) &=& \frac{1}{\Gamma\left(\frac{1}{q -
1}\right)}\frac{T}{q - 1}\left(\frac{1}{q -
1}\frac{T}{\tilde{T}}\right)^{\frac{2 - q}{q - 1}}\cdot \exp\left(
- \frac{1}{q - 1}\frac{T}{\tilde{T}}\right)
\label{eq:gamma}\\
&& {\rm where}\qquad q-1=\frac{Var(1/\tilde{T})}{\langle
1/\tilde{T}\rangle^2} .\label{eq:defq}
\end{eqnarray}
Convoluting $\exp (- E/\tilde{T})$ with such a $f(1/\tilde{T})$
one obtains immediately Tsallis distribution $h_q(E)$
(\ref{eq:Tsallis}) \cite{WW}. The parameter $q$, i.e., according
to Eq. (\ref{eq:defq}) also the temperature fluctuation pattern,
is therefore fully given by the parameters describing this basic
diffusion process (cf., \cite{WW} for details) \footnote{This was
recently generalized to account for the possibility of
transferring energy from/to a heat bath, which appears to be
important for AA applications \cite{WW_epja,WWprc} and for cosmic
ray physics \cite{WWcosmic}; however, we shall not discuss this issue here).}.\\

The second point is that, as was shown in \cite{fluct},
temperature fluctuations in the form given by Eq. (\ref{eq:gamma})
result in an automatic broadening of the corresponding
multiplicity distributions, $P(N)$, from the poissonian form for
the exponential distributions, Eq. (\ref{eq:BG}),
\begin{equation}
P(N) = \frac{\left( \bar{N}\right)^N}{N!} \exp\left ( -
\bar{N}\right) \quad {\rm where}\quad \bar{N} =\frac{E}{T}.
\label{eq:Poisson}
\end{equation}
to the negative binomial (NB) form for the Tsallis distributions,
Eq. (\ref{eq:Tsallis}) (cf., \cite{fluct}, for details),
\begin{equation}
P(N)\, =\, \frac{\Gamma(N+k)}{\Gamma(N+1)\Gamma(k)}\frac{\left(
\frac{\langle N\rangle}{k}\right)^N}{\left( 1 + \frac{\langle
N\rangle}{k}\right)^{(N+k)}};\quad {\rm where}\quad
k=\frac{1}{q-1}.\label{eq:NBD}
\end{equation}
Notice that in the limiting cases of $q\rightarrow 1$ one has
$k\rightarrow \infty$ and (\ref{eq:NBD}) becomes a poissonian
distribution (\ref{eq:Poisson}), whereas for $q\rightarrow 2$ on
has $k\rightarrow 1$ and (\ref{eq:NBD}) becomes a geometrical
distribution. It is easy to show that for large values of $N$ and
$\langle N\rangle$ one obtains from Eq. (\ref{eq:NBD}) its scaling
form,
\begin{equation}
\langle N\rangle P(N) \cong  \psi\left( z=\frac{N}{\langle
N\rangle} \right) = \frac{k^k}{\Gamma(k)} z^{k-1}\exp( - kz),
\label{eq:scalingform}
\end{equation}
in which one recognizes a particular expression of
Koba-Nielsen-Olesen (KNO) scaling \cite{KNO} \footnote{It is worth
mentioning at this point that, as shown in \cite{RWW},
fluctuations of $\bar{N}$ in the poissonian distribution
(\ref{eq:Poisson}) taken in the form of $\psi(\bar{N}/<N>)$, Eq.
(\ref{eq:scalingform}), lead to the NB distribution
(\ref{eq:NBD}).}.\\

\section{Results}
\label{section:II}

Proceed to a detailed analysis of different observables, like
$P(N)$, $dN/dy$ or $dN/dp_T$, fluctuations in which are expected
to differ from each other and therefore to result in different
values of the corresponding parameters $q$. Indeed, from our
experience with $p{\bar p}$ collisions \cite{compq} we know that
one can obtain a very good description of the whole range of $p_T$
spectra ($\propto \exp_q\left( -p_T/T \right)$ with $(T = T_T$
[GeV]; $q_T)= (0.134;1.095)$, $(0.135;1.105)$ and $(0.14;1.11)$
for energies (in GeV) $200$, $540$ and $900$, respectively
\footnote{There is recent compilation of essentially all results
for $p_T$ spectra, including recent LHC data \cite{Wibig}. It
shows that $q(s) = 1.25 - 0.33 s^{-0.054}$, which nicely
reproduces results mentioned here.}. These values should be
compared with the corresponding values of $(T=T_L;q=q_L)$ obtained
when fitting rapidity distributions ($\propto \exp_q\left( -\mu_T
\cosh y/T\right)$) at the same energies: $(11.74;1.2)$,
$(20.39;1.26)$ and $(30.79;1.29)$. It was noticed there that $q_L
-1$ has the same energy behavior as $1/k$ in the NB distribution
fitting the multiplicity distributions at corresponding energies
($q_L - 1 = -0.104 + 0.058 \ln \sqrt{s}$). This means that
fluctuations of the total energy are in this case mainly driven by
fluctuations in longitudinal phase space. An explanation proposed
in \cite{compq} was the following. Noticing that $q-1 =
\sigma^2(T)/T^2$ (i.e., it is given by fluctuations of total
temperature $T$) and assuming that $\sigma^2(T) = \sigma^2(T_L) +
\sigma^2(T_T)$, one can estimate that the resulting values of $q$
should not be too different from
\begin{equation}
q\, =\, \frac{q_L\, T_L^2\, +\, q_T\, T^2_T}{T^2}\, -\,
        \frac{T^2_L\, +\, T^2_T}{T^2}\, +\, 1 \quad \stackrel{T_L \gg
        T_T}{\Longrightarrow}\quad \sim q_L.
\label{eq:qqq}
\end{equation}

\begin{figure}[t]
\includegraphics[width=0.5\textwidth]{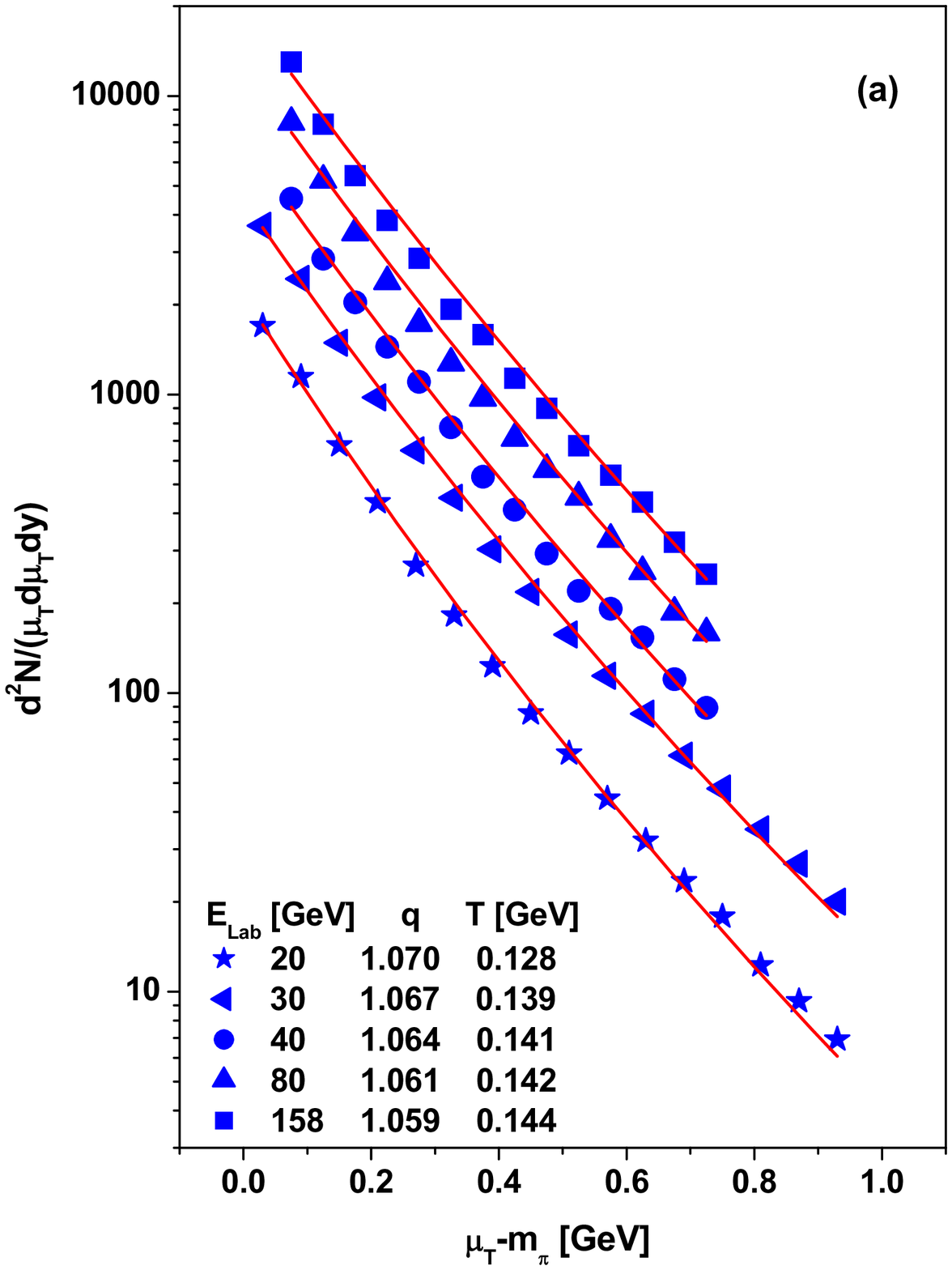}
\includegraphics[width=0.5\textwidth]{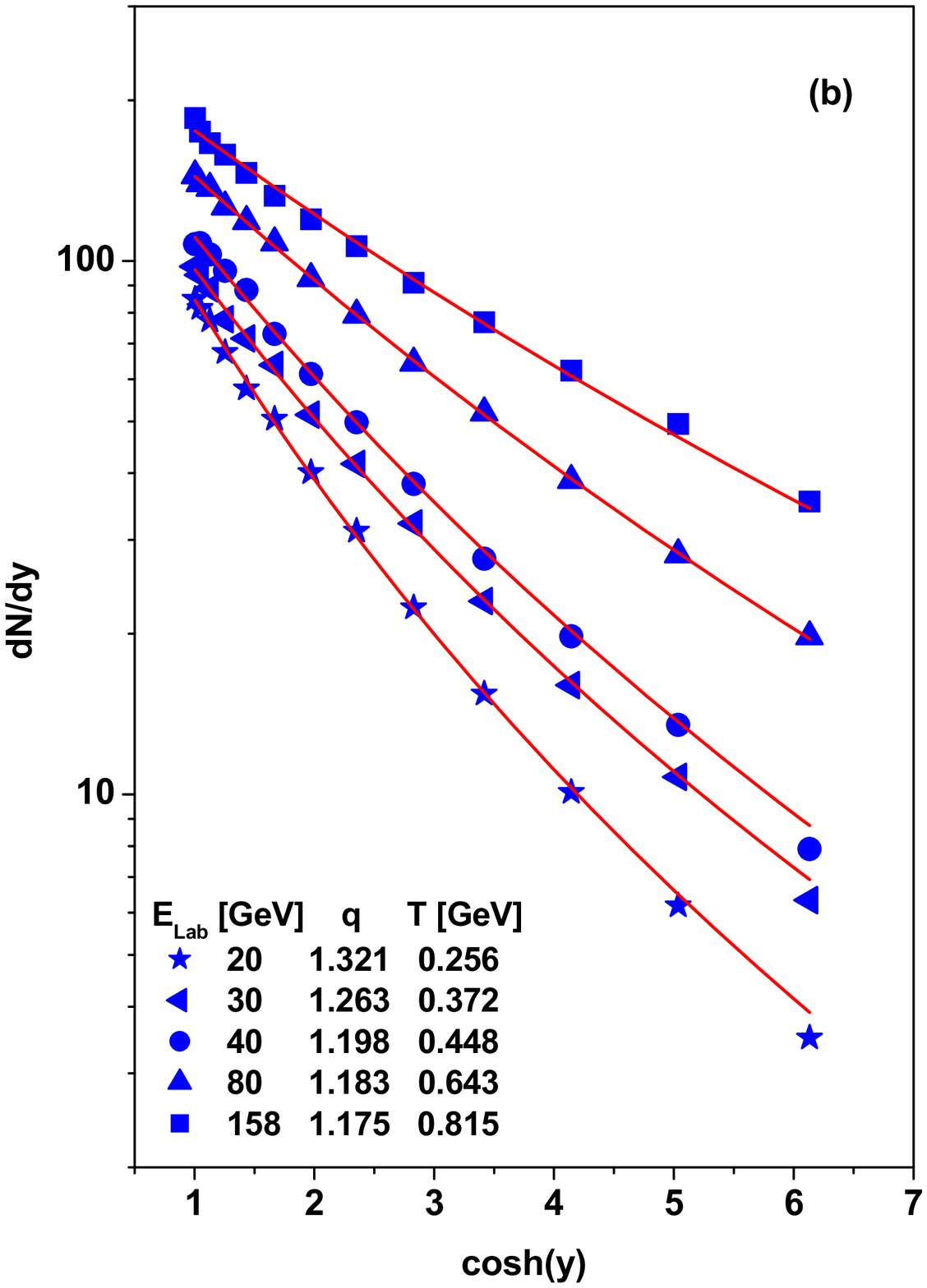}
\caption{(Color online) Fits using Eq. (\ref{eq:Tsallis}) to
\cite{NA49} data for $dN/d\mu_T$ (a) and for $dN/dy$ (b) for
central collision $Pb+Pb$ at different energies.}\label{Fig1}
\end{figure}

\begin{figure}[t]
\includegraphics[width=0.5\textwidth]{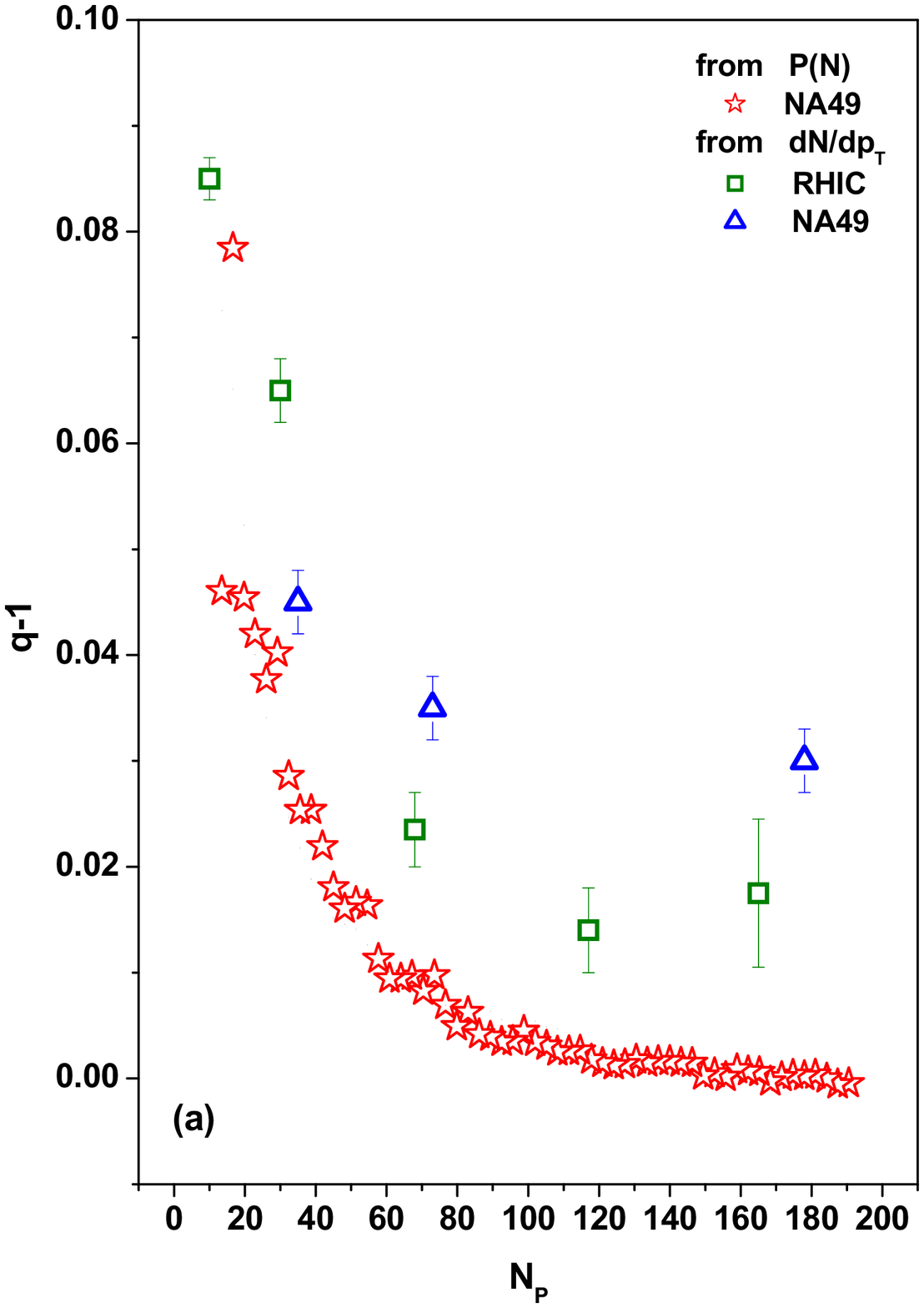}
\includegraphics[width=0.5\textwidth]{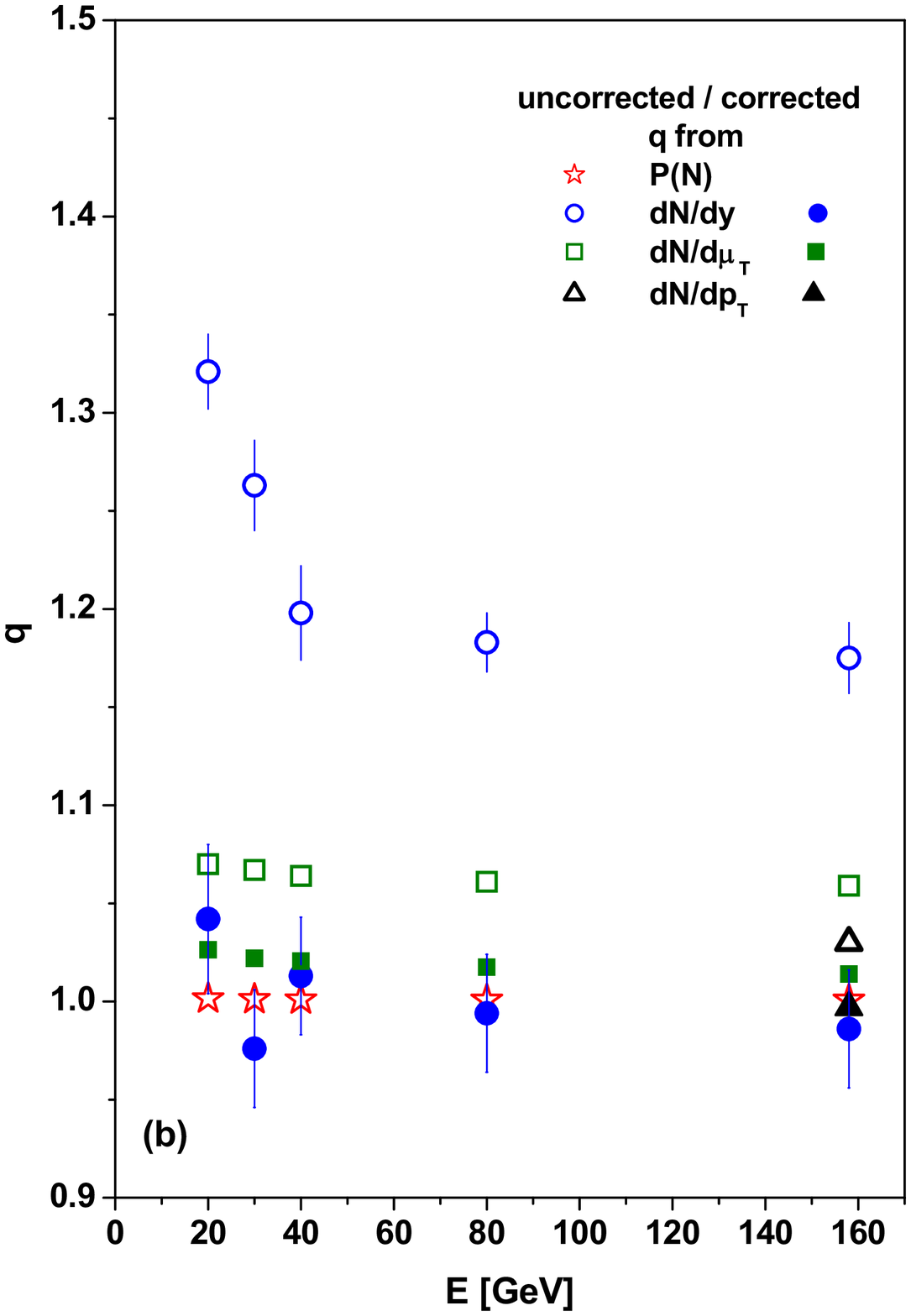}
\caption{(Color online) $(a)$: $q$ for different centralities
measured by the number of projectile participants, $N_P$. Here $q$
from $P(N)$ are our results from $Var(N)/\langle N\rangle$
\cite{WWprc} to be compared with $q=q_T$ obtained from $dN/dp_T$
data of NA49 (cf. first work of \cite{NA49}). PHENIX results
\cite{RHIC} analyzed in \cite{qcompilation} are also shown for
comparison. $(b)$:  Energy dependence of $q$ for the most central
events. All results were obtained for the sake of this
presentation using distributions provided by \cite{NA49}, i.e.,
respectively, $dN/d\mu_T$, $dN/dy$ and $dN/dp_T$. The errors are
similar to those presented as an example for $q = q_L$ obtained
from $dN/dy$. Open symbols correspond to uncorrected values of
$q$, full symbols to values corrected by means of the procedure
proposed in this work and explained in the text.}\label{Fig2}
\end{figure}

The situation is noticeably different for nuclear collisions,
which we shall now address\footnote{For this purpose we use mainly
NA49 data on $Pb+Pb$ collisions \cite{NA49} because, at the
moment, only this experiment measures both (at least for the most
central collisions) multiplicity distributions, $P(N)$, and
distributions in rapidity $y$, transverse momenta, $p_T$, and
transverse masses, $\mu_T = \left( m^2 + p^2_T \right)^{1/2}$, and
this property is crucial for further considerations. PHENIX
results \cite{RHIC} analyzed in \cite{qcompilation} are also shown
for comparison.}. As shown in Fig. \ref{Fig1}, data for $dN/dy$
and $dN/d\mu_T$ can be fitted perfectly by means of Eq.
(\ref{eq:Tsallis}). However, the behavior of the $q$ parameters
obtained is quite interesting, as displayed in Fig. \ref{Fig2}. At
first, the parameter $q$ from $P(N)$ turns out to depend on the
centrality of the collision defined by the number of participants
of projectile, $N_P$ (left panel of Fig. \ref{Fig2}),
\begin{equation}
q - 1 = \frac{1}{aN_P}\left( 1 - \frac{N_P}{A} \right)
\label{eq:qP(N)}
\end{equation}
($A$ - mass number of colliding nuclei and $a = 0.98$)
\cite{WWprc}. Whereas for small centralities it approaches
situation encountered in $p\bar{p}$, the more central the event,
the smaller is $q -1$, i.e., the nearer to a poissonian the
corresponding $P(N)$.
This time for each centrality $q_T$ are larger than $q$ (both for
results based on NA49 data \cite{NA49} and from fits presented in
\cite{qcompilation} based on PHENIX data \cite{RHIC}). In right
panel of Fig. \ref{Fig2} we collected all results for the most
central events from NA49 \cite{NA49}. Notice that they clearly
display opposite trend to that encountered for the $p\bar{p}$
collisions mentioned above: both, $q_L$ and $q_T$ (obtained from
$p_T$ distributions are now greater than $q$ and have
(approximately) a visible similar dependence on $N_P$. However
now $q_L < q_T$, again, this is opposite to what was seen in $p\bar{p}$.\\

A natural question is, what causes such different behavior of the
parameter $q$ in this case? The answer we propose: When extracting
values of $q$ from the rapidity distributions a tacit assumption
was that $\mu_T$ in $E = \mu_T\cosh y$ remains constant (i.e., it
does not fluctuate). However, this is too crude, because data show
that $\mu_T$ fluctuates as well. To account for this fact notice
that $\exp_q\left( -E/T\right) = \exp_q\left[ -
\left(\mu_T/T\right)\cosh y\right] = \exp_q( - z\cosh y)$, i.e.,
that fits to rapidity distributions provide us with fluctuations
not so much of partition temperature $T$ but rather of the
variable $z = \mu_T/T$. This in turn can be written approximately
as:
\begin{equation}
Var(z) \simeq \frac{1}{\langle T\rangle^2}Var\left( \mu_T\right) +
\frac{\langle \mu_T \rangle^2}{\langle T\rangle^2} \cdot
\frac{Var(T)}{\langle T\rangle^2}. \label{eq:varz}
\end{equation}
Because
\begin{equation}
\langle z\rangle \simeq \frac{\langle \mu_T\rangle}{\langle
T\rangle} \quad \rm{and}\quad \frac{Var(1/T)}{\langle
1/T\rangle^2} \simeq  \frac{Var(T)}{\langle T\rangle^2}
\label{eq:der1}
\end{equation}
and because
\begin{equation}
\frac{Var(z)}{\langle z\rangle^2} =  \frac{Var\left(
\mu_T\right)}{\langle \mu_T \rangle^2} + \frac{Var(T)}{\langle
T\rangle^2} \label{eq:der2}
\end{equation}
one obtains following sum rule connecting different fluctuations
\begin{equation}
q - 1 \stackrel{def}{=} \frac{Var(T)}{\langle T\rangle^2} =
\frac{Var(z)}{\langle z\rangle^2} - \frac{Var\left(
\mu_T\right)}{\langle \mu_T\rangle^2} . \label{eq:sumrule}
\end{equation}

This sum rule is our main result and its action is presented in
the right panel of Fig. \ref{Fig2}. It connects total $q$, which
can be obtained from the analysis of the NB form of the measured
multiplicity distributions, P(N), with $q_L -1 = Var(z)/\langle
z\rangle^2$, obtained from fitting rapidity distributions and
$Var\left(\mu_T\right)/\langle \mu_T\rangle^2$ obtained from data
on transverse mass distributions. When extracting $q$ from
distributions of $dN/d\mu_T$ we proceed analogously with
$z=\cosh y/T$.\\

\begin{figure}[h]
\includegraphics[width=0.5\textwidth]{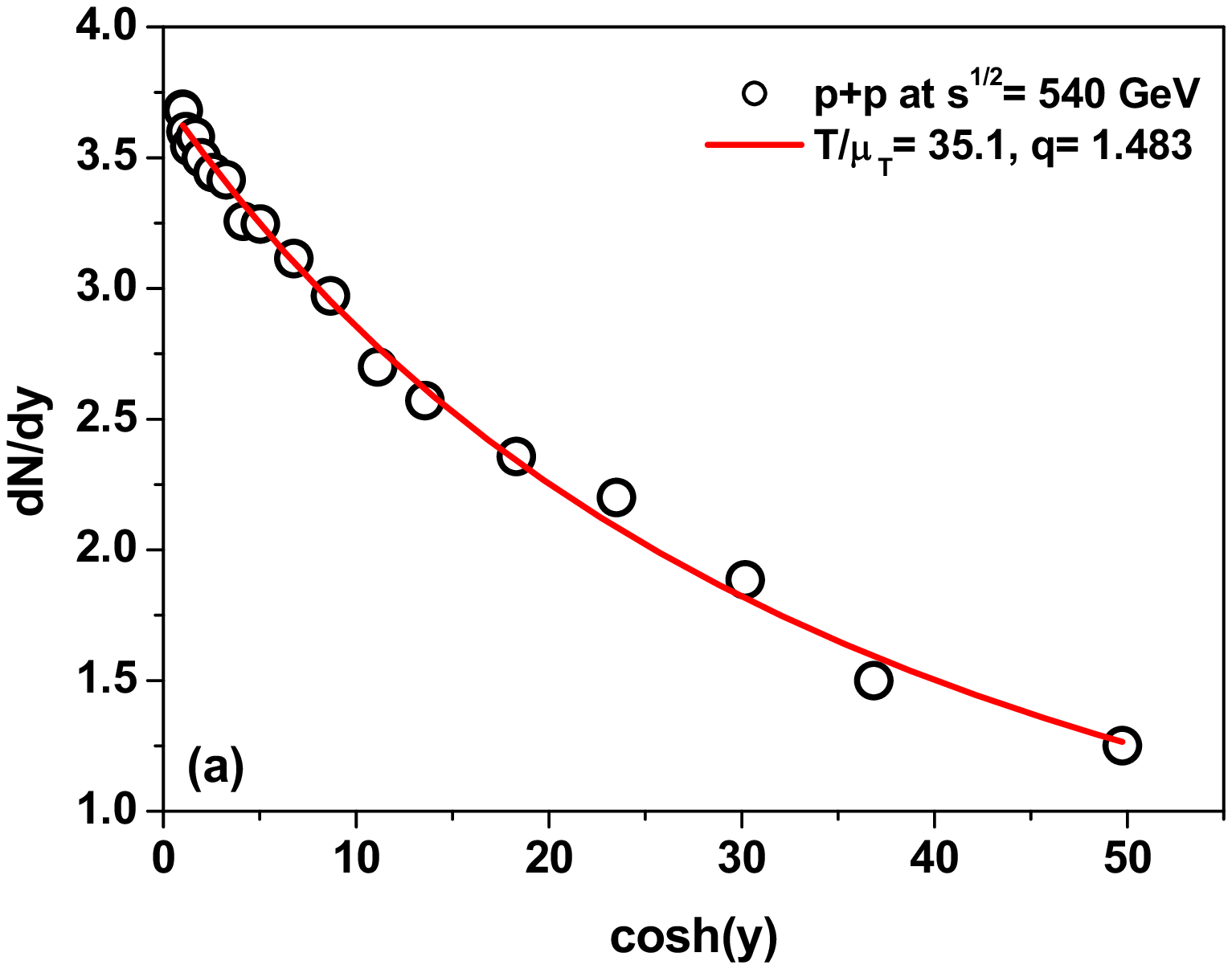}
\includegraphics[width=0.5\textwidth]{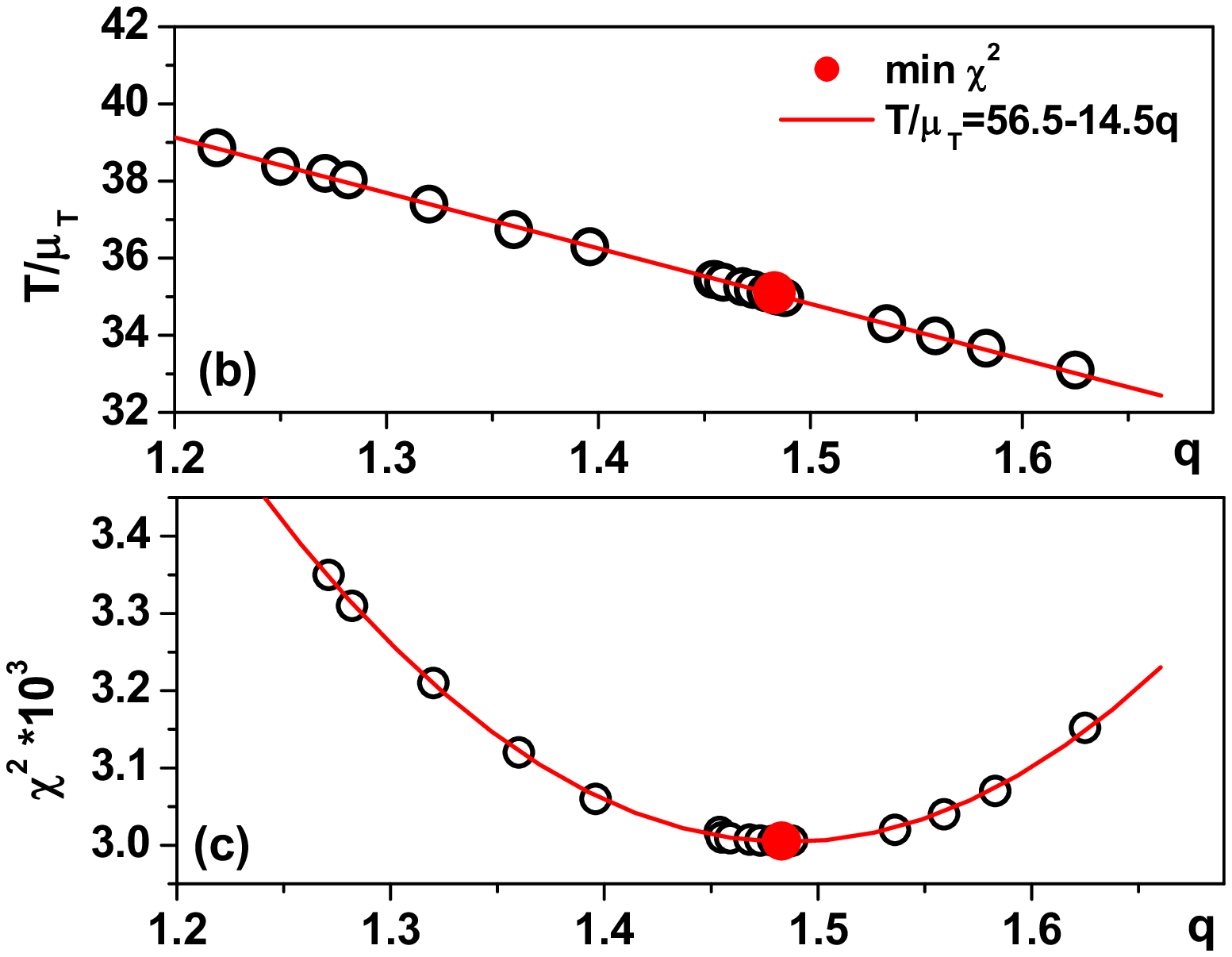}
\caption{(Color online) $(a)$: Example of best fit to $dN/dy$ for
$E_{CM}=540$ GeV (data are the same as in \cite{RWW}). $(b)$: The
$q$-dependence of the admissible $T/\mu_T$. $(c)$: $q$-dependence
of the corresponding $\chi^2$. The full red dot shows the minimal
value of $\chi^2$, see text for details. }\label{Fig3}
\end{figure}

Some explanatory remarks on the apparent discrepancies between
$pp$ and $AA$ data are in order here. When fitting $dN/dy$ data on
$pp$ in \cite{RWW} $\mu_T$ was kept constant (and given by the
$\langle p_T\rangle$ for given energy). This means that all
effects related to its fluctuations was attributed to the
fluctuations of the "partition temperature" $T=T_L$ (which is
therefore only one of the fitted parameters and, for example, it
cannot be used to calculate the mean energy). This was possible
because, as observed in \cite{qcompilation,PHENIX}, in the fitting
procedure parameters $T$ and $q$ are strongly correlated. To
illustrate this fact we present in Fig. \ref{Fig3} an example of
the best fit to $dN/dy$ for $E_{CM} = 540$ GeV together with
$q$-dependency of $T/\mu_T$ and parameter $\chi^2$ representing
the goodness of the fit\footnote{Note that here $\chi^2$ is not
exactly the same as that commonly used in statistics. Namely, for
$m$ experimental values $x_i$ compared with values of $x_i'$ given
by the fitting formula with $m'$ parameters we use $\chi^2 =
\sum_{i=1}^m\left(x_i - x_i'\right)^2/\left(m - m'\right)$.}.
Notice that for different values of $q$ and $T/\mu_T$ we can
describe $dN/dy$ with reasonable accuracy. Parameters $T$ and $q$
are strongly correlated, for $E_{CM} = 540$ GeV shown here
$T(q)/\mu_T = 56.5  - 14.5 q$. Comparing values of $T/\mu_T$ and
$q$ shown in Fig. \ref{Fig3} to those reported for this energy
earlier \cite{RWW} ($q' = 1.26$ and $T'/\mu_T = 35.1$) one can
observe that the difference $\Delta q = q - q' = 0.22$ is roughly
the same as the correction caused by fluctuations of $\mu_T$
discussed above\footnote{We have chosen for our analysis nuclear
collision data from NA49 because all of them (for different
energies) are coming from the same experiment, i.e., they were
obtained under the same experimental conditions. Because of this
fact comparison of different $q$ (i.e., different fluctuations)
was relatively simple. To repeat this procedure for the $pp$ data
would be much more complicated and involved and demands a separate
investigation.}.\\

\section{Summary}
\label{section:III}

We have demonstrated that fluctuations of temperature $T$,
together with fluctuations of other variables, result in the sum
rule formula, Eq. (\ref{eq:sumrule}), connecting $q$ obtained from
an analysis of different distributions. This allows us to
understand why in $AA$ collisions fluctuations observed in
multiplicity distributions are much smaller than the corresponding
ones seen in the rapidity distribution or in the distribution of
transverse momenta (i.e., why the corresponding $q$ parameters
evaluated from distributions of different observables are
different). This issue should be checked further when complete
sets of data become available from the experiments at LHC
(especially from ALICE).

\section*{Acknowledgements}
Partial support (GW) of the Ministry of Science and Higher
Education under contract DPN/N97/CERN/2009 is acknowledged.

\end{document}